\documentclass{article}
\usepackage{spconf,amsmath,graphicx}
\usepackage{multirow}
\usepackage{caption}
\usepackage[para,online,flushleft]{threeparttable}

\makeatletter
\newcommand{\thickhline}{%
    \noalign {\ifnum 0=`}\fi \hrule height 1pt
    \futurelet \reserved@a \@xhline
}

\newcommand{\comment}[1]{}

\title{NON-AUTOREGRESSIVE END-TO-END APPROACHES FOR JOINT AUTOMATIC SPEECH RECOGNITION AND SPOKEN LANGUAGE UNDERSTANDING}
\name{Mohan Li and Rama Doddipatla}
\address{Toshiba Europe Ltd, Cambridge Research Laboratory}
\copyrightnotice{978-1-6654-7189-3/22/\$31.00~\copyright2023 IEEE}
\begin{document}
%
\maketitle
\begin{abstract}
This paper presents the use of non-autoregressive (NAR) approaches for joint automatic speech recognition (ASR) and spoken language understanding (SLU) tasks. The proposed NAR systems employ a Conformer encoder that applies connectionist temporal classification (CTC) to transcribe the speech utterance into raw ASR hypotheses, which are further refined with a bidirectional encoder representations from Transformers (BERT)-like decoder. In the meantime, the intent and slot labels of the utterance are predicted simultaneously using the same decoder. Both Mask-CTC and self-conditioned CTC (SC-CTC) approaches are explored for this study. Experiments conducted on the SLURP dataset show that the proposed SC-Mask-CTC NAR system achieves 3.7\% and 3.2\% absolute gains in SLU metrics and a competitive level of ASR accuracy, when compared to a Conformer-Transformer based autoregressive (AR) model. Additionally, the NAR systems achieve 6$\times$ faster decoding speed than the AR baseline.
\end{abstract}
\section{Introduction}
\label{sec:intro}

Automatic speech recognition (ASR) has been serving as the interface between human users and a series of downstream natural language understanding (NLU) tasks, such as intent classification (IC) and slot filling (SF) \cite{tur2011spoken}. Traditionally, the ASR and NLU modules are cascaded to construct a spoken language understanding (SLU) system \cite{mesnil2014using,coucke2018snips}, where the spoken utterance is first transcribed into text and from which semantic arguments are extracted. However, since the components of the traditional system are typically optimised using separate criteria, the ASR errors may severely degrade the following NLU performance, especially under noisy conditions. Furthermore, the execution time of such a pipeline architecture could also become an obstacle for its real-world deployment, where quick system response is often required.

With the recent advances of neural network, there is growing popularity of designing SLU systems in the end-to-end (E2E) fashion \cite{serdyuk2018towards,saxon2021end,haghani2018audio}, where the ASR and NLU components are integrated into a single network and optimised with a joint loss function. The E2E SLU systems generally adopt the encoder-decoder-based sequence-to-sequence (Seq2Seq) framework, which has been widely employed in several areas including neural machine translation \cite{sutskever2014sequence,bahdanau2015neural} and ASR \cite{bahdanau2016end,chan2016listen}. The encoder of the system transforms the input (audio or text) to higher level representations, and then the decoder maps them to the corresponding output. According to \cite{haghani2018audio}, the architecture of E2E SLU can be categorised as follows: (1) direct model, which learns semantic arguments directly from the input audio without transcribing the full utterance; (2) joint model, whose decoder jointly predicts ASR transcripts and semantic arguments; (3) multitask model, where the ASR and SLU tasks are performed using separate decoders but share a common acoustic encoder; (4) multistage model, which resembles the traditional pipeline system in having two sets of encoder-decoder structures for ASR and NLU respectively, while the two modules are connected by propagating the gradients through both stages.

Among the aforementioned model types, the direct model, which exclusively rules out ASR transcription, delivers less promising SLU performance than the others, particularly on the SF task. This reveals that ASR plays a crucial role in predicting semantic arguments like the slot labels, which makes the component indispensable to E2E SLU systems. For the rest of the model structures, the joint model has shown to achieve competitive SLU scores with the most compact architecture, making it suitable for on-device applications. While in the multitask and multistage models, since both the ASR and SLU decoders perform language modelling, it is reasonable to amalgamate them into an unified network, which in return facilitates the tasks to benefit each other with shared knowledge. Thus in this work, we aim to establish joint ASR and SLU systems that consist of a single encoder-decoder framework.

So far, most of the joint SLU systems have been implemented as autoregressive (AR) models, represented by the well-known Transformer architecture \cite{vaswani2017attention}. A prominent shortcoming of the AR models is that the decoder can only predict one output at a time, which significantly increases the execution time when the utterance length grows. By contrast, the non-autoregressive (NAR) models have much faster inference speed as the entire output sequence is simultaneously generated. Recently, a number of NAR models \cite{higuchi2021comparative} have been proposed for ASR tasks, with the majority using the connectionist temporal classification (CTC) technique as the backbone to fulfill parallel decoding. However, as CTC does not account for the dependencies between output tokens, the vanilla NAR ASR models usually tend to achieve inferior performance when compared to their AR counterparts.

To relax the conditional independence assumption adopted by CTC-based models, Mask-CTC \cite{higuchi2020mask} incorporates a language model (LM) decoder that is trained to predict masked tokens within the input text. At test time, the low-confidence ASR outputs produced by the CTC encoder can be iteratively refined through their reliance on the other confident ones. Besides, instead of utilising a decoder, intermediate-CTC \cite{lee2021intermediate} involves an iterated CTC loss that is calculated on the outputs of some middle encoder layers. By introducing such a regularisation strategy, the linguistic contexts are implicitly augmented to the lower layers so as to support the subsequent layers' predictions. Self-conditioned CTC (SC-CTC) \cite{nozaki2021relaxing} further extends the idea by feeding back the intermediate ASR posterior at certain encoder layers to the input of the next layer. Compared with intermediate CTC, SC-CTC provides more explicit linguistic biases to the acoustic encoder, and the reliability of those grows as it approaches the top layer. Other NAR algorithms such as Improved Mask-CTC \cite{higuchi2021improved}, Imputer \cite{chan2020imputer} and Align-Denoise \cite{chen2021align} are also proved effective in closing the performance gap with AR models.

Motivated by the above NAR ASR models, this paper proposes two joint ASR and SLU systems that build upon Mask-CTC and SC-CTC approaches, respectively. The encoder of both systems applies the Conformer architecture \cite{gulati2020conformer}. And to enhance the SLU performance, we implement the decoder as a bidirectional encoder representations from Transformers (BERT)-like structure \cite{devlin2019bert}, which predicts the IC and SF targets at the same time \cite{chen2019bert}. Meanwhile, the raw ASR hypotheses produced by the encoder are refined using the same decoder. The proposed systems were verified on an open-sourced dataset called SLURP \cite{bastianelli2020slurp}. In the experiments, both systems showed to outperform the Transformer-based AR model regarding IC accuracy and SF SLU-F1 scores, and achieve competitive ASR performance. The proposed NAR SLU systems also realised noticeably faster inference when compared with the AR baseline. 

The rest of the paper is organised as follows: Section 2 presents the basis of NAR ASR approaches as Mask-CTC and SC-CTC. Section 3 describes the proposed joint SLU systems based on the above NAR methods. Experimental details on the SLURP dataset are demonstrated in Section 4. Finally, conclusions are drawn in Section 5.

\section{Non-autoregressive ASR Models}
\label{sec:nar_asr}

NAR models have gained increasing popularity within the ASR community following their wide success on neural machine translation tasks \cite{gu2017non,libovicky2018end}. In contrast with the AR models which incrementally predict the output sequence, NAR ASR enables transcribing all the text tokens concurrently across the utterance. Such a decoding manner undoubtedly brings about higher inference speed, hence making the NAR models preferable in many real-time interaction scenes. In this section, we'll review two prevailing NAR ASR methods that are easily adopted to perform joint E2E ASR and SLU.

\subsection{Mask-CTC}
\label{ssec: mask_ctc}

Mask-CTC has been proposed to mitigate the problem of a pure CTC-based model that the dependencies between output tokens are principally neglected. In this regard, Mask-CTC uses a conditional masked language model (CMLM) \cite{ghazvininejad2019mask} based decoder to progressively refine the primitive ASR hypotheses produced by the original CTC encoder. 

During training, the encoder is still supervised with standard CTC loss $\mathcal{L}_{ctc}$, while the CMLM decoder is made to predict the randomly masked tokens $\mathbf{Y}_{mask}$ within the ground-truth sequence $\mathbf{Y}$, conditioning on the remaining observed ones $\mathbf{Y}_{obs}$ and the representations $\mathbf{X}$ that are the outputs of the final encoder layer:
\begin{equation} 
    P_{cmlm}(\mathbf{Y}_{mask} | \mathbf{Y}_{obs}, \mathbf{X}) = \prod_{y \in \mathbf{Y}_{mask}} P(y | \mathbf{Y}_{obs}, \mathbf{X}).
\label{eq:p_cmlm}
\end{equation}
The objective of the CMLM decoder is thus computed as:
\begin{equation} 
    \mathcal{L}_{cmlm} = - \mathrm{log} \; P_{cmlm}(\mathbf{Y}_{mask} | \mathbf{Y}_{obs}, \mathbf{X}).
\label{eq:loss_cmlm}
\end{equation}
Finally, the overall loss function of the Mask-CTC model is given as a weighted sum of the CTC and CMLM parts:
\begin{equation} 
    \mathcal{L}_{mask-ctc} = \lambda \mathcal{L}_{ctc} + (1 - \lambda) \mathcal{L}_{cmlm},
\label{eq:loss_mask_ctc}
\end{equation}
where $\lambda$ is a hyperparameter tuned on the development set.

During inference, the best path CTC alignment $\mathbf{\hat{a}}_{ctc}$ is first obtained through greedy search decoding on $\mathbf{X}$, and subsequently collapsed to the raw ASR hypotheses sequence $\mathbf{\hat{Y}}^0$ by removing the blank and repetition symbols. Then starting from $\mathbf{\hat{Y}}^0$, the CMLM decoder consecutively refines the ASR transcripts for a maximum of $M$ iterations. At the $i^{th}$ iteration, the output tokens in $\mathbf{\hat{Y}}^{i-1}$ that have insufficient confidence scores $\hat{P}$ are replaced with the $\langle \mathrm{MASK} \rangle$ symbol:
\begin{equation} 
    \mathbf{\hat{Y}}^{i-1}_{mask} = \{ y \in \mathbf{\hat{Y}}^{i-1} | \hat{P}(y) < P_{thresh} \},
\label{eq:mask_y}
\end{equation}
where $P_{thresh}$ denotes the threshold that determines whether a token should be masked out. Thereby, as in training $\mathbf{\hat{Y}}^{i-1}_{mask}$ are updated by the CMLM decoder conditioning on the unmasked tokens $\mathbf{\hat{Y}}^{i-1}_{obs}$ and encoder representations $X$, to produce the refined transcripts $\mathbf{\hat{Y}}^i$. The above process repeats until either all the tokens in a certain $\mathbf{\hat{Y}}$ exceed the threshold, or the maximum number of iterations are reached.

Thanks to the refinement operations based on mask prediction, Mask-CTC is able to outperform standard CTC by a large margin, while retaining the competent decoding speed. However, Mask-CTC suffers from the mismatch between training and inference stages, since the CMLM decoder is optimised on the ground-truth transcripts but works with the ASR hypotheses during inference. For instance, the insertion errors may greatly affect the CMLM performance at inference time, as these are never seen within the training data.

\subsection{Self-conditioned CTC}
\label{ssec: sc_ctc}

Unlike Mask-CTC that relies on an additional LM decoder, SC-CTC handles the conditional independence issue completely inside the encoder architecture itself. At a certain encoder layer $l$, the probability distributions over the vocabulary are calculated for all encoding timesteps from the output $\mathbf{X}^l_{out}$, which are referred to as the intermediate prediction $\mathbf{Z}^l$. It is obvious that the linguistic information contained in $\mathbf{Z}^l$ could be used to condition the encoding of following layers. Therefore, $\mathbf{Z}^l$ is added onto $\mathbf{X}^l_{out}$ to augment the input of the next encoder layer:
\begin{equation} 
    \mathbf{X}^{l+1}_{in} = \mathrm{LayerNorm} (\mathbf{X}^l_{out}) + \mathrm{Linear}_{|V| \rightarrow D} (\mathbf{Z}^l),
\label{eq:in_sc_ctc}
\end{equation}
where $\mathrm{LayerNorm(\cdot)}$ stands for layer normalisation \cite{ba2016layer} and $\mathrm{Linear}_{|V| \rightarrow D} (\cdot)$ means linear projection that transforms $\mathbf{Z}^l$ from the vocabulary size $|V|$ to the model dimension $D$. The above operation can be carried out on multiple middle layers (usually at equal intervals), and as it proceeds to the top layer, the quality of $\mathbf{Z}^l$ is gradually improved to provide more reliable contextual dependencies to the encoding. Besides, the parameters of the linear projection are shared among all the intermediate layers, where no significant increase of model size is introduced to the original architecture. 

Similar to intermediate CTC, the objective of SC-CTC consists of the CTC losses from both the final encoder layer and the $K$ intermediate layers:
\begin{equation} 
    \mathcal{L}_{sc-ctc} = \eta \mathcal{L}_{ctc} + (1 - \eta) \mathcal{L}_{inter-ctc},
\label{eq:loss_sc_ctc}
\end{equation}
\begin{equation} 
    \mathcal{L}_{inter-ctc} = - \frac{1} {K} \sum^K_{l=1} \mathrm{log} \; P_{ctc} (\mathbf{Z}^l).
\label{eq:loss_iter}
\end{equation}
where $\eta$ denotes a parameter used to weigh the CTC losses.

Compared with Mask-CTC, SC-CTC features various advantages in the following aspects: (1) Simpler architecture, as neither additional decoder nor major modification to the standard CTC encoder is required; (2) Unified training and inference processes, since the entire system only takes spoken utterances as input at both stages; (3) More efficient linguistic augmentation, where the encoding is directly conditioned on the latent ASR outputs predicted at several intermediate layers. While in the Mask-CTC based model, the decoder can only add contextual dependence to the ASR hypotheses as the final output of the encoder.

\section{Proposed NAR SLU Systems}
\label{sec:nar_slu}

Inspired by the NAR ASR algorithms presented above, in this section, we propose two E2E SLU systems that build on the corresponding NAR framework. Both systems are supposed to jointly perform ASR and two major SLU tasks, IC and SF. 

\subsection{Mask-CTC SLU}
\label{ssec: mask_ctc_slu}

The Mask-CTC architecture is believed naturally applicable to E2E SLU, given that the CMLM decoder shares the similar fundamental logic with a NLU module, with both taking ASR transcripts as input and conducting semantic feature extractions. To facilitate efficient SLU, in our proposed system we carry out the CMLM decoding in the fashion of a BERT model, so that the intent and slot labels are predicted simultaneously in a single output sequence.

When running the BERT model, a special symbol $\langle \mathrm{CLS} \rangle$ is inserted prior to the input sequence, whose output embedding is assumed holding the sentence-level representation. Therefore, during training we prepend the partially masked ground-truth transcripts with $\langle \mathrm{CLS} \rangle$ to get the input $\mathbf{Y}$ to the CMLM decoder. In this case, the decoder adopts a separate output head for the SLU arguments, where the first hidden state $\mathbf{h}_0$ of CMLM (corresponds to $\langle \mathrm{CLS} \rangle$) is used to detect the intent $o_i$ and the rest ones $\{\mathbf{h}_1,...,\mathbf{h}_N\}$ are used to predict the slot labels $\mathbf{o}_s = \{ o^1_s, ..., o^N_s \}$:
\begin{equation} 
    P_{cmlm}(\mathbf{O} | \mathbf{Y}, \mathbf{X}) = P(o_i | \mathbf{h}_0) \prod^N_{n=1} P(o^n_s | \mathbf{h}_n),
\label{eq:p_icsf}
\end{equation}
where $\mathbf{O} = [ o_i; \mathbf{o}_s ]$, and $N$ is the length of the ground-truth transcripts. Meanwhile, the masked tokens in $\mathbf{Y}$ are predicted in the same way as Mask-CTC ASR following eq. (\ref{eq:p_cmlm}).

The objective of the CMLM decoder is turned into two folds, given as:
\begin{equation} 
    \begin{aligned}
    \mathcal{L}_{joint-cmlm} &= \gamma \mathcal{L}^{ASR}_{cmlm} + (1 - \gamma) \mathcal{L}^{SLU}_{cmlm}\\
    &= - \gamma \mathrm{log} \; P_{cmlm}(\mathbf{Y}_{mask} | \mathbf{Y}_{obs}, \mathbf{X})\\ 
    &\quad - (1 - \gamma) \mathrm{log} \; P_{cmlm}(\mathbf{O} | \mathbf{Y}, \mathbf{X}),
    \end{aligned}
\label{eq:loss_joint_cmlm}
\end{equation}
where $\gamma$ is the weight to balance the ASR and SLU tasks. Finally, the overall loss function of the Mask-CTC SLU system is computed as a sum of the CTC and the extended CMLM terms like in eq. (\ref{eq:loss_mask_ctc}).

Likewise, at inference time $\langle \mathrm{CLS} \rangle$ is added before the ASR hypotheses sequence $\mathbf{\hat{Y}}$ at each refinement iteration, where the IC-SF predictions $\mathbf{\hat{O}}$ are updated together with $\mathbf{\hat{Y}}$ via the CMLM decoder.

\begin{figure*}[t]
\centering
  \centerline{\includegraphics[width=12cm]{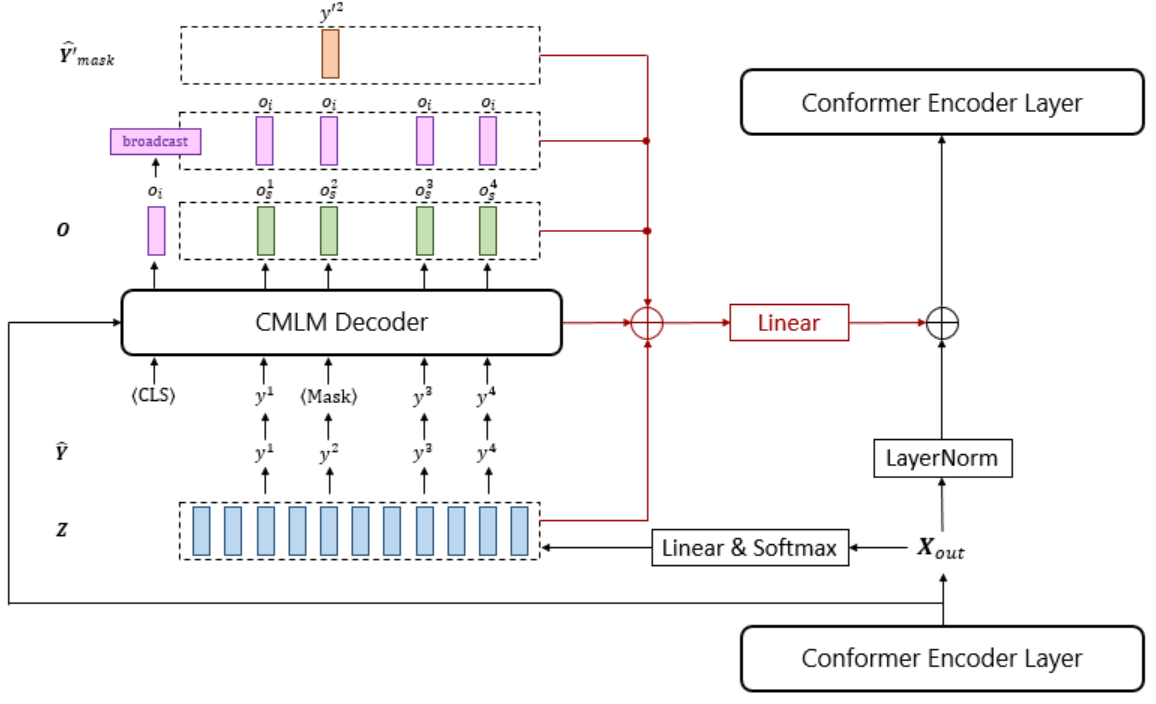}}
\caption{The workflow of SC-Mask-CTC decoding at an intermediate encoder layer.}
\label{fig:sc_mask_ctc}
%
\end{figure*}

\subsection{SC-Mask-CTC SLU}
\label{ssec: sc_mask_ctc_slu}

Although SC-CTC could achieve solid ASR performance only with a pure CTC-based encoder, such an architecture is deemed inadequate for implementing SLU systems. This is because the extraction of semantic features heavily relies on the contextual dependence within the utterance, which conflicts with the CTC's assumption. As a result, we propose an approach called SC-Mask-CTC, to equip the encoder with the aforementioned BERT-like CMLM, such that most of the advantages of the original SC-CTC algorithm are preserved.

As an extension to the SC-CTC ASR system, the CMLM decoder only operates on the encoder layers where ASR prediction is conducted, i.e. the intermediate layers and the final layer. For the intermediate layers, taking layer $l$ for example, since the input to CMLM is a sequence of discrete linguistic units (e.g. word-pieces), we need to derive the ASR hypotheses $\mathbf{\hat{Y}}^l$ from the probability distribution $\mathbf{Z}^l$ via CTC greedy search decoding. Concurrently, the positions on the CTC alignment $\mathbf{\hat{a}}^l_{ctc}$ at which the tokens in $\mathbf{\hat{Y}}^l$ are selected will also be recorded as $\mathbf{S}^l = \{s^l_1, .., s^l_N\}$ for later use. Note that when collapsing the repetition within $\mathbf{\hat{a}}_{ctc}$, the position with the most confident output token is chosen into $\mathbf{S}^l$.

Similar to the inference stage of Mask-CTC SLU, we first mask out the low-confidence tokens in $\mathbf{\hat{Y}}^l$ and place the $\langle \mathrm{CLS} \rangle$ symbol to its front. The resulted sequence is then fed to the CMLM decoder to jointly predict the masked tokens $\mathbf{\hat{Y}}^l_{mask}$, intent $o^l_i$ and slot labels $\mathbf{o}^l_s$.

Apart from the intermediate prediction $\mathbf{Z}^l$, we seek to further augment the next layer's input with all the outcomes of CMLM, namely the refined ASR tokens $\mathbf{\hat{Y'}}^l_{mask}$ and the SLU arguments $\mathbf{O}^l$. It should be noticed that $\mathbf{\hat{Y'}}^l_{mask}$ and $\mathbf{O}^l$ have distinctive lengths and both are much shorter than $\mathbf{Z}^l$. Thus, for $\mathbf{\hat{Y'}}^l_{mask}$ and $\mathbf{o}^l_s$, we add their probability distributions to the specific positions of $\mathbf{Z}^l$ according to $\mathbf{S}^l$. While for $o^l_i$, the probability distribution is broadcast-add to the positions of $\mathbf{Z}^l$ following $\mathbf{S}^l$ as a sentence-level bias term. To summarise, the input to the next encoder layer is computed as:
\begin{equation}
    \begin{aligned}
    \mathbf{X}^{l+1}_{in} &= \mathrm{LayerNorm} (\mathbf{X}^l_{out})\\
    &\quad + \mathrm{Linear}_{|V| \rightarrow D} (\mathbf{Z}^l \oplus [\mathbf{\hat{Y}}^l_{mask}, \mathbf{O}^l | \mathbf{S}^l]),
    \end{aligned}
\label{eq:in_sc_ctc_slu}
\end{equation}
where $\oplus$ denotes the adding strategy proposed above. An illustration of the above workflow at a certain intermediate layer is provided in Fig. \ref{fig:sc_mask_ctc}

As for the final encoder layer, the inference of the ASR transcript $\mathbf{\hat{Y}}$ and the SLU arguments $\mathbf{\hat{O}}$ is just a single iteration of CMLM decoding as done in the intermediate layers. Consequently, $\mathbf{\hat{Y}}$ and $\mathbf{\hat{O}}$ are used as final output of the system.

The objective of SC-Mask-CTC is defined as the combination of the SC-CTC and joint CMLM losses given in eq. (\ref{eq:loss_sc_ctc}) and (\ref{eq:loss_joint_cmlm}), respectively:
\begin{equation} 
    \mathcal{L}_{sc-mask-ctc} = \mu \mathcal{L}_{sc-ctc} + (1 - \mu) \mathcal{L}_{joint-cmlm},
\label{eq:loss_sc_ctc_slu}
\end{equation}
where $\mu$ denotes a hyperparameter to regulate the losses. 

During training, the CMLM outputs at the intermediate layers are not included in the computation of $\mathcal{L}_{joint-cmlm}$, instead the CMLM is only trained with the ground-truth ASR transcripts as in the Mask-CTC approach. This inevitably creates mismatch between training and inference. However, one could observe that such inconsistency only falls on the final encoder layer, while the decoding manner at all the intermediate layers remains the exactly same in both stages.

Moreover, we believe the linguistic augmentation that SC-Mask-CTC brings to the encoding is even stronger than in vanilla SC-CTC, as it not only includes the intermediate predictions at various encoder layers, but also the CMLM refined ASR transcripts, as well as the inferred SLU labels which provides semantic-level knowledge.

\begin{table*}
\centering
\caption{An example of the ASR transcripts and SLU annotations in the SLURP dataset.}
\begin{tabular}{lllllllll}
\thickhline
\multicolumn{1}{l}{\bf{ASR transcripts}} & 
& \multicolumn{1}{c}{set}
& \multicolumn{1}{c}{lunch}
& \multicolumn{1}{c}{every}
& \multicolumn{1}{c}{day}
& \multicolumn{1}{c}{at}
& \multicolumn{1}{c}{twelve}
& \multicolumn{1}{c}{thirty}
\\
\hline
\multicolumn{1}{l}{\bf{Slot labels}} & 
& \multicolumn{1}{c}{O}
& \multicolumn{1}{c}{B\_meal\_type}
& \multicolumn{1}{c}{B\_general\_frequency}
& \multicolumn{1}{c}{I\_general\_frequency}
& \multicolumn{1}{c}{O}
& \multicolumn{1}{c}{B\_time}
& \multicolumn{1}{c}{I\_time}
\\
\multicolumn{1}{l}{\bf{Intent}} & & \multicolumn{7}{c}{calendar\_set} \\
\thickhline
\end{tabular}
\label{tab:example}
\end{table*}

\section{Experiments}
\label{sec:exp}

\subsection{AR SLU baseline}
\label{ssec: ar_baseline}

The AR baseline to our proposed NAR SLU systems adopts the joint CTC/attention \cite{kim2017joint} based Conformer-Transformer (C-T) architecture that is popular within the ASR community. At decoding step $j$, the baseline model predicts the an ASR token $y^j$ and a SLU argument $o^j$ using separate output heads. As an AR decoding strategy, the predictions of $y^j$ and $o^j$ are conditioned on all the previously decoded ASR tokens $\{y^1, .. y^{j-1}\}$ that have been step-wise fed back to the decoder. The decoding proceeds until an end-of-sequence symbol $\langle \mathrm{EOS} \rangle$ is generated from the ASR head at step $L$. Then we obtain the final ASR transcripts as $\mathbf{\hat{Y}} = \{y^1, ..., y^{L-1}\}$, and the aligned slot labels as $\mathbf{o}_s = \{o^1, ..., o^{L-1}\}$. The last SLU output $o^L$ that emerges with $\langle \mathrm{EOS} \rangle$ is selected as the intent $o_i$, since it is predicted by seeing all the preceding ASR tokens to access sentence-level information. Beam-search decoding is carried out for the AR baseline in terms of ASR outputs, meanwhile only the most probable SLU output is kept for each decoding step. No external LM is utilised in the decoding process.

\subsection{Experimental setup}
\label{ssec: exp_setup}

The proposed NAR SLU systems have been evaluated on the SLURP dataset, which contains 83 hours, more than 140k spoken utterances from the domain of in-home personal robot assistant. All experiments are conducted using ESPnet toolkit \cite{watanabe2019}. The acoustic feature of the audio comprises 80-dimensional filterbanks along with 3-dimensional pitch-related parameters, with SpecAugment \cite{park2019specaugment} applied to the training set. The vocabulary of the systems combines (a) 500 BPE \cite{sennrich016neural} tokenised word-pieces for ASR transcription, (b) 70 intent classes, (c) 56 slot labels, prepended with 'B\_' and 'I\_' to indicate the position of the word in the slot filler, as well as (d) 5 functional symbols such as $\langle blank \rangle$ and $\langle CLS \rangle$. As a result, the total size of the vocabulary sums up to 682. An example of the ASR transcripts and the corresponding SLU annotations is provided in Table. \ref{tab:example}.

All the systems are equipped with a 2-layer convolutional neural network (CNN) based frond-end, which adopts the kernel size of 3 and a stride of 2 that reduces the frame-rate by 4 times altogether. The Conformer encoder has 12 layers, where for each layer the CNN kernel size, attention dimension, number of heads and feed-forward network size are \{15, 256, 4, 2048\}. The same architecture is employed for the Transformer decoder of the baseline system and the CMLM decoder of the proposed NAR systems, which has 6 layers and shares identical network parameters with the encoder. The total number of parameters in the system is around 44M. We use 3 intermediate layers as the \{$3^{rd}$, $6^{th}$, $9^{th}$\} layer in the SC-Mask-CTC encoder. For the NAR losses, the weights \{$\lambda$, $\eta$, $\gamma$, $\mu$\} are set to \{0.4, 0.5, 0.5, 0.4\}. And the CTC weight of the AR system is set to 0.3 for both training and inference processes. The ASR confidence threshold $P_{thresh}$ is chosen as 0.999 for Mask-CTC, and \{0.9, 0.99, 0.999, 0.999\} for the intermediate and final encoder layers of the SC-Mask-CTC model in the down-to-top order. The baseline, Mask-CTC and SC-Mask-CTC SLU systems are trained up to \{100, 400, 400\} epochs with the mini-batch size of 64, since the NAR models usually take more data iterations to converge when compared to the AR models. To achieve the best balance between inference speed and performance, we set the beam-width of AR decoding to 5, and the maximum number of refinement iterations $M$ of Mask-CTC to 10. All the decoding jobs are conducted on Intel(R) Xeon(R) Gold 5220S CPUs.

\subsection{Experimental results}
\label{ssec: exp_re}

\begin{table}[t]
\centering
\caption{Word error rates (WERs \%), IC accuracy (Acc \%), SF SLU-F1 score (\%) and real-time factor (RTF) achieved by joint ASR and SLU systems on the test set of SLURP.}
\begin{threeparttable}
\begin{tabular}{lllll}
\hline \thickhline
\multicolumn{1}{l}{Model}  & \multicolumn{1}{c}{WER}  &  \multicolumn{1}{c}{IC Acc}  & \multicolumn{1}{c}{SF SLU-F1}  & \multicolumn{1}{c}{RTF} \\ \thickhline \hline
\multicolumn{3}{l}{AR}   \\ \hline
\multicolumn{1}{l}{C-T (ours)}   & \multicolumn{1}{c}{15.7}     & \multicolumn{1}{c}{85.4}  & \multicolumn{1}{c}{74.3}  & \multicolumn{1}{c}{0.52}  \\
\multicolumn{1}{l}{RNNT-BERT \cite{Raju2022}}   & \multicolumn{1}{c}{15.2}  & \multicolumn{1}{c}{82.5}      & \multicolumn{1}{c}{72.4}  & \multicolumn{1}{c}{-} \\
\multicolumn{1}{l}{Transformer \cite{omachi2022non}}  & \multicolumn{1}{c}{*20.5}   & \multicolumn{1}{c}{-}      & \multicolumn{1}{c}{74.3}  & \multicolumn{1}{c}{2.27}       \\ \thickhline \hline
\multicolumn{3}{l}{NAR}      \\ 
\hline
\multicolumn{1}{l}{g-Mask-CTC \cite{omachi2022non}}     & \multicolumn{1}{c}{*25.1}  & \multicolumn{1}{c}{-}      & \multicolumn{1}{c}{68.8}  & \multicolumn{1}{c}{0.15}  \\
\multicolumn{1}{l}{t-Mask-CTC \cite{omachi2022non}}     & \multicolumn{1}{c}{*25.2}  & \multicolumn{1}{c}{-}      & \multicolumn{1}{c}{68.7}  & \multicolumn{1}{c}{0.16}   \\
\multicolumn{1}{l}{\bf{Mask-CTC}}   & \multicolumn{1}{c}{\bf{17.2}}  & \multicolumn{1}{c}{\bf{88.5}}      & \multicolumn{1}{c}{\bf{77.2}}  & \multicolumn{1}{c}{\bf{0.08}}  \\
\multicolumn{1}{l}{\bf{SC-Mask-CTC}}   & \multicolumn{1}{c}{\bf{16.8}}  & \multicolumn{1}{c}{\bf{89.1}}      & \multicolumn{1}{c}{\bf{77.5}}  & \multicolumn{1}{c}{\bf{0.09}}  \\
\hline \thickhline
\end{tabular}
\begin{tablenotes}
\item The numbers with * are estimated from a figure in the original paper where the exact values were not provided.
\end{tablenotes}
\end{threeparttable}
\label{tab:results}
\end{table}

Table. \ref{tab:results} presents the ASR and SLU performance achieved by the baseline AR and proposed NAR SLU systems, together with other reference systems in literature. It can be observed that regarding the ASR performance, the AR systems consistently obtain lower word error rates (WERs) than the NAR counterparts. However, unlike the Mask-CTC based systems in \cite{omachi2022non} that greatly fall behind their AR baseline, our proposed NAR systems achieve relatively close ASR accuracy to the AR system. Specifically, the SC-Mask-CTC model outperforms the Mask-CTC model with an absolute gain of 0.4\% WER, showing the advantage of applying the CMLM decoder between the encoder layers rather than on the top.

The SLU performance is evaluated using the prediction accuracy (Acc) and SLU-F1 score \cite{bastianelli2020slurp} for the IC and SF tasks, respectively. The IC Acc is defined as the proportion of the utterances with correctly detected intent in the whole test set. The SF SLU-F1 score extends the standard F1 score by taking the ASR errors into consideration. The edit distance between the golden and predicted slot fillers is derived on both word and character levels, which are used as penalty terms in the F1 score computation. From Table. \ref{tab:results}, we could see that the proposed NAR SLU systems achieve significantly better results on both IC and SF tasks, when compared to the AR baseline and all the reference systems in literature. The Mask-CTC model outperforms the AR baseline with absolute gains of 3.1\% in IC Acc and 2.9\% in SF SLU-F1 score. While the gains achieved by the SC-Mask-CTC model are even 3.7\% and 3.2\% for the two metrics, delivering the state-of-the-art SLU performance at 89.1\% IC Acc and 77.5\% SF SLU-F1 score on the SLURP dataset. We attribute the capability of the propose systems to the BERT-like CMLM decoder, which enables the capture of bidirectional context information and the joint prediction of intent and slot labels. Whereas these features are completely incompatible with the decoding fashion of the AR system.

The inference speed of the SLU systems is measured with real-time factor (RTF), a ratio of the execution time to the length of the input utterance. As expected, the proposed NAR systems achieve much lower RTF (around 6$\times$) than the AR baseline, indicating higher inference speed. One may also notice that Mask-CTC yields a slightly better RTF compared to SC-Mask-CTC. Through investigation we find that the CMLM decoder of the Mask-CTC model runs for average 2.5 iterations on the test set of SLURP, which is smaller than the fixed $K+1=4$ turns of decoding in SC-Mask-CTC.


\section{Conclusions}
\label{sec:con}

In this paper, we propose to use NAR approaches, known as Mask-CTC and SC-Mask-CTC, for E2E joint ASR and SLU. The systems based on both approaches build upon the encoder-decoder framework, where the encoder adopts the Conformer architecture and the decoder is implemented as a CMLM that decodes in the fashion of a BERT model. In the Mask-CTC based system, the raw ASR hypotheses are first generated by the encoder via CTC greedy search decoding, in which the low-confidence ASR tokens are replaced with a mask symbol. Then the resulted sequence is input to the CMLM decoder to iteratively refine the masked tokens, and at the same time predict the intent and slot labels simultaneously. On the other hand, the SC-Mask-CTC based system conducts the similar CMLM decoding at a range of intermediate encoder layers, where the refined ASR transcripts and predicted SLU arguments are further fed back to the model to condition the encoding of following layers. We carried out experiments on the SLURP dataset, and the results show that both proposed NAR systems outperformed the Transformer-based AR baseline on the IC and SF tasks. The SC-Mask-CTC model even achieved the start-of-the-art SLU performance on SLURP with 89.1\% IC Acc and 77.5\% SF SLU-F1 score. Also, the NAR systems obtained competitive ASR accuracy relative to the AR system. With respect to the inference speed, the NAR approaches yielded approximately 6$\times$ faster decoding when compared to the AR method.

\bibliographystyle{IEEEbib}
\bibliography{refs}

\end{document}